\documentstyle[11pt]{article}

\newlength{\bibwidth}        
\def\al           {\alpha}

\def\de           {\delta}
\def\ep           {\epsilon}

\def\ga           {\gamma}

\def\la           {\lambda}

\def\si           {\sigma}

\def\De           {\Delta}
\def\Ga           {\Gamma}
\def\La           {\Lambda}

\def\Pi           {\Pi}

\def\non          {\nonumber}

\def\ha           {\mbox{$\frac{1}{2}$}}
\def\thi          {\mbox{$\frac{1}{3}$}}
\def\qu           {\mbox{$\frac{1}{4}$}}

\def\pl           {\partial}
\def\d        {\mbox{d}}
\def\spa          {\ \ \ }
\def\mand         {\spa\mbox{and}\spa}
\def\diag	  {\mbox{diag}}

\def\cd           {{\cdot}}
\def\ran          {\rangle}
\def\lan          {\langle}

\def\M		  {T}

\newcounter {sect}
\def\thesect {\arabic{sect}}
\def\sect#1{
\setcounter{subsect}{0}
\refstepcounter{sect}
\noindent{\large{\bf \thesect. #1}}
}

\newcounter {subsect}
\def\thesubsect {\arabic{subsect}}
\def\subsect#1{
\refstepcounter{subsect}
\noindent{\large{\thesect.\thesubsect. {\em #1}}}\newline
}

\newcounter {inbib}
\def\theinbib {\arabic{inbib}}
\def\inbib#1#2{
\refstepcounter{inbib}
{\begin{minipage}[t]{7mm}{\begin{flushright}\theinbib.
\end{flushright}}\end{minipage} 
\parbox[t]{\bibwidth}{#2}
\label{#1}\newline}
}

\def\bib#1{$^{\mbox{\scriptsize{\ref{#1}}}}$}

\scrollmode

\begin{document}

\pagenumbering{arabic}
\hfill hep-th/9804012
$\mbox{}$

\vspace{3cm}
\begin{center}
{\Large{\bf MASSIVE FIELDS AND THE 2D STRING}}

\vspace{1cm}
\renewcommand{\thefootnote}{\fnsymbol{footnote}}
Andy Wilkins\footnote{E-mail: awilkins@physics.adelaide.edu.au} 
\setcounter{footnote}{0}

{\em Department of Physics and Mathematical Physics \\
University of Adelaide \\
Adelaide 5005 \\
Australia}

(1 April 1998)

\vspace{5mm}

\begin{minipage}{14cm}
The first massive level of closed bosonic string theory is studied.
Free-field equations are derived by imposing Weyl invariance on the
world sheet.  A two-parameter solution to the equation of motion and
constraints is found in two dimensions with a flat linear-dilaton
background.  One-to-one tachyon scattering is studied in this
background.  The results support Dhar, Mandal and Wadia's  proposal
that 2D critical string theory corresponds to the $c=1$ matrix model
in which both sides of the Fermi sea are
excited.
\end{minipage}
\end{center}

\vspace{5cm}
\noindent ADP-98-9/M63

\newpage

\sect{Introduction}
\label{sec.intro}

\setcounter{footnote}{0}
\renewcommand{\thefootnote}{\alph{footnote}}

It is well known that the requirement of Weyl and reparameterisation
invariance implies constraints on the spacetime fields of critical
string theory (see Green, Schwarz and Witten\bib{gsw} and the
references contained therein).  These constraints govern the dynamics
of the spacetime fields and so can be used to predict the outcome of
various scattering experiments.

For the bosonic string in an empty flat background, cancellation of
the Weyl anomaly implies a critical dimension of 26, a fact which was
first elucidated by Polyakov\bib{poly}.  Weyl invariance with a
background graviton field was first studied by
Friedan\bib{friedan}.  One of the achievements of his thesis was the
calculation the graviton's beta-function to 
two-loops using a normal coordinate expansion in the partition function.
Setting the beta-function to zero, up to 
a  curl, is equivalent to demanding Weyl invariance.  The result is
that the spacetime metric must be Ricci flat (to 
one loop, up to a curl) in order that the theory have vanishing Weyl
anomaly.  Callan et.\ al~\bib{cfmp} found the coupled field equations
for all massless fields in the bosonic string by setting 
the beta functions equal to zero.  Their
calculation has since been extended beyond one loop.  The tachyon
field can also be added without ruining the renormalisability
of the theory.  However, Das and Sathiapalan\bib{ds} noticed that
with the 
inclusion of the tachyon, there were contributions to the Weyl anomaly
that were invisible to any finite order in the loop-expansion.  These
contributions can be obtained by using a weak-field expansion
instead of the loop expansion.  Since
the goal of this paper is to study tachyon scattering, a weak field
expansion will have to be employed. 

A popular method of implementing the weak-field expansion is the
``Wilson renormalisation group'' approach which was pioneered by Banks 
and Martinec\bib{bm} and developed by Hughes et.\ al\bib{jjj}.  All
the massive levels of the string are included which makes the theory
non-renormalisable.  The theory is regulated with a short-distance
cutoff, and scale 
invariance is imposed by asserting that 
the couplings must be at a fixed point of their Wilson
renormalisation-group flows.  This paper studies the dynamics of the
first massive level within the Wilson renormalisation group framework.
The free field equations are obtained at linear order in the
expansion.  At the quadratic order, the situation becomes vastly more 
complicated as, generically, every field will contribute to every
equation of motion.  Therefore, it must be assumed that it is
consistent to study a finite subset of string modes by taking the
other modes to be higher-order in this expansion.  This point will be
returned to when the 
coupled field equations are derived in Sec.~\ref{sec.method}.

String theory in two target-space dimensions is both non-trivial and
solvable.  With a linear dilaton background there is only one
propagating particle, it is massless and is conventionally called the
`tachyon'.  The higher `discrete modes' are non-propagating.  The
first such is a blackhole solution for the 2D metric-dilaton system.
Blackhole physics is naturally non-perturbative and thus beyond the
reach of the 
purely perturbative description of string theory discussed so far.  2D
string theory is exciting because it has a well-known non-perturbative
description through its identification with the non-critical $c=1$
string.  This non-perturbative description is the double-scaled $c=1$
matrix model (see the  
reviews by Ginsparg and Moore\bib{gm} and Polchinski\bib{polchrev}
and the references therein).  In 
principle then, here is a toy model in which blackhole evolution and
the stringy effects on black holes can be studied.

In the double scaling limit, the $c=1$ matrix model consists simply of
nonrelativistic, free fermions living in an inverted harmonic
oscillator potential.  The 
Fermi surface lies just below the top of the 
potential.  Once an identification between excitations in the matrix
model and the spacetime fields is found, the free-fermion picture
can be used to calculate the result of any spacetime scattering
experiment.  It has been shown\bib{mp} that the spacetime 
tachyons are related through the `leg-pole transform' to the
bosonised fluctuations of the Fermi surface.  This identification was
made possible by comparing bulk scattering amplitudes of 
Louiville field theory\bib{difk} with small-pulse
scattering in the matrix model\bib{polch}.   The leg-pole transform
is nonlocal and it gives rise to the spacetime gravitational physics
which is absent in the free-fermion picture.  The discrete modes were
not identified until relatively recently.

Dhar, Mandal and Wadia\bib{dmw}$^{,}$\bib{dmw2}
(DMW) considered fluctuations of the Fermi
surface on both sides of the potential.  Until their paper, working
with both sides of the Fermi sea was deemed unnecessary for
small pulses in the semi-classical region, since tunnelling is a
non-perturbative occurrence.  However, considering both sides gave two
scalars, the average and the difference of the bosonised fluctuations
on each side of the sea.  The authors compared scattering in this
model with the effective tachyon-graviton theory (with a linear
dilaton) obtained by imposing Weyl invariance.  It was shown that the
tachyon was the leg-pole transform of the average, 
while the mass of the blackhole was the energy associated with
the difference.  They postulated that the other discrete modes 
corresponded to higher moments of the difference variable.  This could
not be checked since the effective theory including the higher modes
had not yet been worked out.

The aim of this paper is to check the proposition of DMW by
considering the first massive mode of the string.  Sec.~\ref{sec.method}
presents the method that is used to impose Weyl 
invariance of the theory in arbitrary spacetime dimensions.  By way of
example, the tachyonic and massless levels are examined.  In
Sec.~\ref{sec.mass} the first 
massive level at linear order is studied.  The non-propagating
solution to the 
equation of motion and constraints in two dimensions with a flat
linear-dilaton background is presented.  Finally, the effective theory
of tachyons interacting with the massive background is derived.   This
is compared with DMW's prediction in
Sec.~\ref{sec.dmw}.

\sect{Method and an Example}
\label{sec.method}

Consider the closed bosonic string theory defined by the partition function
\begin{equation}
Z = \int \frac{[\d g_{ab}\d\xi]_{g}}{V_{\mbox{\scriptsize{diff}} \times
\mbox{\scriptsize{Weyl}}}} e^{-S[\xi^{\mu},g_{ab}]} \ , 
\end{equation}
where $\xi^{\mu}$ are a collection of $D$ scalar fields living on the
string worldsheet which has the topology of a sphere and metric
${g}_{ab}$.  Latin letters $a,b,\ldots = 1,2$ are used to indicate
worldsheet indices, while the spacetime
indices will always be denoted by Greek 
letters, $\mu = 0,\ldots,D-1$.  Worldsheet reparameterisation
invariance will be kept manifest throughout the calculation, so after
fixing the conformal gauge 
$g_{ab} = e^{2\si}\de_{ab}$ in two patches on the sphere, the
partition function reads\bib{poly}
\begin{equation}
Z = \int \frac{[\d\si]}{V_{\mbox{\scriptsize{Weyl}}}}\left\{ \int
[\d^{D}\xi]_{\de} \exp
\left( \frac{D-26}{24\pi}\int\si\Box\si - S[\xi,g] \right) \right\} \ ,
\end{equation}
with $\Box = \de^{ab}\pl_{a}\pl_{b}$.  Weyl invariance of the
theory means that 
arbitrary correlation functions $\lan \prod_{i}\xi^{\mu_{i}} \ran$
calculated with the path integral contained in the curly parentheses are be
independent of $\si$.  Then the measure $\int
[\d\si]/V_{\mbox{\scriptsize{Weyl}}}$ can be set to unity.  In the
Wilson renormalisation group approach, the equations of motion are
usually obtained as an operator statement inside the path integral.  In
contrast, this
paper utilises a source $J^{\mu}$ and 
explicitly calculates the generating functional
\begin{equation}
Z[J] = e^{\frac{D-26}{24\pi}\int\si\Box\si} \int[\d^{D}\xi]
e^{-S[\xi^{\mu},{g}_{ab}] + \int J\cd f(\xi)} \ ,
\end{equation}
by employing a weak-field expansion and using
a short-distance cutoff.  The coupling $\int J\cd f(\xi)$ will be
explained soon.  This approach is closely related to the one used by
Brustein, Nemeschansky and Yankielowicz\bib{bny}. 

\subsect{The linearised field equations for the tachyon and the
massless fields}
There are a number of subtleties inherent in this method
and these are most easily illustrated by considering the familiar
scenario of a string living in a background of three massless
spacetime fields, the graviton
$G_{\mu\nu}$, the antisymmetric tensor $B_{\mu\nu}$ and the dilaton
$\Phi$, and one tachyonic field $T$.  The action is given by
\begin{equation}
S[\xi,\si] = \frac{1}{4\pi} \int_{\cal M} \left( 
\ha\sqrt{g}
G_{\mu\nu}(\xi)\pl_{a}\xi^{\mu}\pl_{b}\xi^{\nu}g^{ab}
+ \ha B_{\mu\nu}(\xi)\pl_{a}\xi^{\mu}\pl_{b}\xi^{\nu}\ep^{ab}
+ \sqrt{g}R\Phi(\xi)
+ \sqrt{g}T(\xi)
\right) \ ,
\label{eqn.massless.action}
\end{equation}
in which string tension $\al'$ has been set to 2 and $g = \det
g_{ab}$.  
It is
well known from 
beta-function results that strings can consistently propagate in a
flat linear-dilaton background
\begin{equation}
G_{\mu\nu} = \eta_{\mu\nu}\ , \spa 
\Phi(\xi) = Q\cd\xi \mand
B_{\mu\nu} = 0 \ ,
\end{equation}
where $3Q^{2} = 26-D$.  The weak-field expansion used in this paper is
about this background,
\begin{equation}
G_{\mu\nu} = \eta_{\mu\nu} + h_{\mu\nu} \mand \Phi = Q\cd\xi + \phi \ ,
\end{equation}
so that the fields $h_{\mu\nu}$, $\phi$ and $B_{\mu\nu}$ are
considered to be $O(\la)$ where, formally, $\la$ is a small parameter.

The guiding principle in  
writing down Eq.~(\ref{eqn.massless.action}) is that it should be the
most general action 
with at most two derivatives that is reparameterisation
invariant both on the worldsheet and in spacetime.  Therefore, the
term $\frac{1}{4\pi} \int \sqrt{g}
\tilde{A}_{\mu}D^{2}\xi^{\mu}$ must also be considered.  Here $D^{2}$
is the covariant Laplacian 
\begin{equation}
D^{2}\xi^{\mu} = g^{ab}D_{a}\pl_{b}\xi^{\mu} = g^{ab}\left(
\pl_{a}\pl_{b}\xi^{\mu} - \Ga^{c}_{ab}(g)\pl_{c}\xi^{\mu} +
\Ga^{\mu}_{\nu\la}(G)\pl_{a}\xi^{\nu}\pl_{b}\xi^{\la} \right) \ .
\end{equation}
The field $\tilde{A}_{\mu}$ makes no
contribution to S-matrix elements since, to $O(\la)$, the on-shell
condition is $D^{2}\xi^{\mu} = 0$.  In the language of this paper, the
equivalent statement is that $\tilde{A}_{\mu}$ can be soaked up by a
field redefinition of $\xi^{\mu}$.  This is because
\begin{equation}
S[\xi^{\mu}] + \frac{1}{4\pi} \int 
\tilde{A}_{\mu}D^{2}\xi^{\mu} = S[\xi^{\mu} - \tilde{A}^{\mu}(\xi)] + 
\frac{1}{4\pi} \int 
\sqrt{g}R Q_{\mu}\tilde{A}^{\mu} \ ,
\end{equation}
to first order in $\la$.  
Then, with the definitions
\begin{equation}
\xi'{}^{\mu} = \xi^{\mu} - \tilde{A}^{\mu}(\xi)  \mand \phi' = \phi +
Q\cd\tilde{A} \ ,
\label{eqn.shift.tildA}
\end{equation}
and the use of the `covariant' measure\footnote{Covariant measures
will be used frequently throughout this work.  They need to be
covariant under spacetime diffeomorphisms, and they must also be local
worldsheet scalars.}
\begin{equation}
\left[\d^{D}\xi\right]_{\mbox{\scriptsize{cov}}} = \left[\d^{D}\xi
\sqrt{\det \left(\ha G_{\mu\nu} - \nabla_{(\mu}\tilde{A}_{\nu)}
\right)} \right] \ ,
\label{eqn.cov.m.m}
\end{equation}
the partition function reads
\begin{equation}
Z = \int [\d^{D}\xi]_{\mbox{\scriptsize{cov}}}e^{-S[\phi;\xi] -
\frac{1}{4\pi} \int  
\tilde{A}_{\mu}D^{2}\xi^{\mu}}
= \int \left[\d^{D}\xi'
\sqrt{\det \ha G_{\mu\nu}(\xi')}\right] e^{-S[\phi';\xi']} \ ,
\end{equation}
to $O(\la)$.  (In the following calculations the prime on the dilaton
will be dropped.)  This 
procedure would have also worked if $\eta_{\mu\nu}$, instead of
$G_{\mu\nu}$,  had been used in the covariant measure.

It is
also useful to note that $\tilde{A}^{\mu}$ could have been absorbed
directly into the metric since the action is invariant under the
transformations
\begin{equation}
\de \tilde{A}_{\mu} = \La_{\mu} \mand 
\de G_{\mu\nu} = \nabla_{\mu}\La_{\nu} + \nabla_{\nu}\La_{\mu}\ ,
\label{eqn.massl.st}
\end{equation}
where $\nabla_{\mu}$ is the covariant spacetime derivative.
This makes $\tilde{A}^{\mu}$ look like a `St\"{u}ckelberg' field --- a
field which is introduced in order that a massive field theory have a
gauge invariance.  $\tilde{A}^{\mu}$ is not a St\"{u}ckelberg
field in the true sense of the term since $G_{\mu\nu}$ is
massless, however.  Such fields will be encountered at the first
massive level and their corresponding gauges will be fixed by setting
the St\"{u}ckelberg fields to zero.

The source $J^{\mu}$ can be coupled to any worldsheet scalar and the theory
will remain invariant under reparameterisations of the worldsheet.  All
choices will break spacetime reparameterisation invariance so the
equations derived by imposing Weyl invariance will be gauge
fixed.  Different couplings $\int J\cd f$ will correspond to different
gauges.  The physics of the theory should not depend on the gauge
choice, but for the purposes of this paper it is convenient to choose
the source term to be
\begin{equation}
\int J_{\mu} (\xi^{\mu} - \si Q^{\mu}) \ .
\label{eqn.coupling}
\end{equation}
There are two reasons for this particular form.  Firstly, it
handles contributions from the linear part of the dilaton field
exactly.  Secondly, using 
the usual coupling $\int J\cd \xi$ and demanding Weyl-invariance of
the one-point function results in the gauge condition
\begin{equation}
0 = \pl_{\mu}\Phi  + \ha \pl_{\mu}\eta^{\mu\nu}h_{\mu\nu} -
\pl_{\nu}h_{\mu\nu} \ ,
\end{equation}
to $O(\la)$.  A flat linear-dilaton background is obviously
inconsistent with this gauge condition.  Of course this background can
be rotated to be compatible with the gauge and, since the other
equations of motion are covariant, S-matrix elements will be
unaffected.  However, this is an unnecessary nuisance.
Eq.~(\ref{eqn.coupling}) is not really all that exotic since it is well
known from spontaneously broken theories that expanding around
different points in configuration space can be advantageous.  Using
this analogy, the choice
$\int J(\xi + \si Q)$ is equivalent to expanding around the true
vacuum, while the coupling $\int J\cd
\xi$ corresponds to expanding around the unstable maxima --- here there
is a non-zero tadpole that runs away into the vacuum.

Completing the square, the generating functional can be written as
\begin{equation}
Z[J] =  P[\si]\int \left[\d^{D}\xi\right]
\exp \left( -\mbox{$\frac{1}{8\pi}$} \int
\pl_{a}\xi^{\mu}\pl_{a}\xi^{\nu}\eta_{\mu\nu}   + 
\mbox{$\frac{1}{4\pi}$}a_{0}\cd \left(J_{0} -
\mbox{$\frac{2Q}{\sqrt{V}}$} \right) -
S_{\mbox{\scriptsize{int}}}(\xi + X) \right) \ ,
\label{gen.fcn.com.sq}
\end{equation}
where
\begin{equation}
P[\si] = \exp \left(\ha \int J\De J - \left(Q^{2} +
\mbox{$\frac{D - 26}{3}$}\right)
\mbox{$\frac{1}{8\pi}$}\int\si\Box\si\right) \ , 
\end{equation}
$a_{0}^{\mu}$ and $J_{0}^{\mu}$ are the zeromodes of $\xi^{\mu}$
and $J^{\mu}/\sqrt{g}$ respectively (so the term $a_{0}\cd (J_{0} +
2Q/\sqrt{V})$ simply 
ensures momentum conservation) and $S_{\mbox{\scriptsize{int}}}$ is 
the action of Eq.~(\ref{eqn.massless.action}) with $G_{\mu\nu}$ and $\Phi$
replaced by the small fields $h_{\mu\nu}$ and $\phi$ respectively.
The `background
field' $X$ is given by 
\begin{equation}
X^{\mu} \equiv X_{0}^{\mu} + Q^{\mu}\si \equiv \int \De J^{\mu} +
Q^{\mu}\si \ ,
\label{eqn.x.}
\end{equation}
and the propagator $\De$ satisfies
\begin{equation}
-\Box\mbox{$\frac{1}{4\pi}$}\De(z,z') = \de(z-z') -
V^{-1}e^{2\si} \ ,
\end{equation}
where $V$ is the volume of the worldsheet.

Performing a weak field
expansion of 
$\exp(-S_{\mbox{\scriptsize{int}}})$  and utilising the Fourier
transform reduces the problem to 
Gaussian integrals.  These integrals are regulated using a
short-distance cutoff $\ep$.
The unregulated propagator, in stereographic coordinates on a sphere
with constant curvature, is
\begin{equation}
\De(z,w) = -\log \left(
\frac{|z-w|^{2}}{(1+z\bar{z})(1+ w\bar{w})} \right) \ .
\end{equation}
The denominator in the logarithm accounts for the 
zeromode on the sphere.  The zeromode enforces momentum conservation
but seems to play no other role in the calculations.
To keep the theory
reparameterisation 
invariant, the regularised 
propagator must satisfy\bib{dhp}$^{,}$\bib{rand}
\begin{eqnarray}
\De_{\ep}(z,z) & = &  \left( 2 \si(z) - \log\ep^{2}
\right) + O(\ep^{2})
\ , \non \\
\left.\frac{\pl}{\pl z^{a}} \De_{\ep}(z,z') \right|_{z' = z} & = &
 \pl_{a} \si(z)  + O(\ep^{2}) \ .
\end{eqnarray}
Finally, the symmetry that allows the absorption of $\tilde{A}^{\mu}$
into the 
graviton in Eq.~(\ref{eqn.massl.st}) can be made to persist at 
the quantum level (to $O(\la)$) if the regularised propagator
satisfies the Leibnitz-like relation
\begin{equation}
D_{z^{a}} \left( \left[ \left. O_{z} O'_{z'} \De_{\ep}(z,z')\right]
\right|_{z=z'} \right)
=
\left. \left[ D_{z^{a}} \left(  O_{z}O'_{z'} \De_{\ep}(z,z')
\right) \right ] \right|_{z=z'} 
 +
 \left. \left[ O_{z} D_{z'{}^{a}} \left( O'_{z'} \De_{\ep}(z,z')
\right) \right]\right|_{z=z'} \ , 
\label{eqn.leib}
\end{equation}
where $O_{z}$ is a linear function of $D_{z}$ and $O'_{z'}$ a linear
function of 
$D_{z'}$.  This can be seen to be necessary by directly calculating
the path integral to $O(\la)$ with $\tilde{A}^{\mu}$ included.  At the
first massive level, the St\"{u}ckelberg fields may be absorbed if
Eq.~(\ref{eqn.leib}) holds 
for quadratic $O$ and $O'$.

Performing the weak-field expansion to $O(\la)$ and using the flat
measure $[\d^{D}\xi]$, the path integral yields
\begin{eqnarray}
Z[J] & = & P[\si] \left( \frac{1}{V} \det{}' \frac{\Box}{4\pi}
\right)^{-\ha D} 
\left\{ \de^{D}(p^{\mu})	
- \frac{1}{4\pi} \int \d^{2}z e^{ipX} e^{-\ha
p^{2}\De_{\ep}(z,z)} \right.
\non \\
&& \spa
\times \left[
 \ep^{-2} e^{2\si} T(p) + \Box \si \phi(p)
\left. +
\ha\left(
\pl_{a}X^{\mu}\pl_{b}X^{\nu} + 2 i p^{\mu} \frac{\pl}{\pl
z_{1}^{a}} \De_{\ep}(z,z_{1}) \pl_{b}X^{\nu} \right)
\right|_{z_{1} = z} \ep^{ab}B_{\mu\nu}(p)
\right.
\non \\
&& \spa\spa
+ \ha \left(
\pl_{a}X^{\mu}\pl_{a}X^{\nu} + 2 i p^{\mu}
 \frac{\pl}{\pl
z_{1}^{a}} \De_{\ep}(z,z_{1}) \pl_{a}X^{\nu} 
+ \eta^{\mu\nu}
 \frac{\pl}{\pl z_{1}^{a}}  \frac{\pl}{\pl z_{2}^{a}} 
\De_{\ep}(z_{1},z_{2}) 
\right.
\non \\
&& \left.\left. \spa\spa\spa\spa
- \left. \left. p^{\mu}p^{\nu} 
\left( \frac{\pl}{\pl
z_{1}^{a}} \De_{\ep}(z,z_{1}) \right)^{2}
\right)\right|_{z_{1}= z_{2} = z} h_{\mu\nu}(p)
\right] \right\} \ .
\label{eqn.first.order}
\end{eqnarray}
In this formula $\det{}'$ is the
determinant without the 
zeromode,
the tachyon has been scaled by $\ep^{-2}$ for convenience and
\begin{equation} 
p^{\mu} + J_{0}^{\mu}\sqrt{V} - 2Q^{\mu} = 0  \ .
\end{equation}
Although the source is a worldsheet density, it must not vary under
Weyl transformations.  Thus $p^{\mu}$ is
Weyl neutral and $\de X^{\mu} = Q^{\mu}\de\si$, as prescribed by
Eq.~(\ref{eqn.x.}).

The second derivatives of $\De_{\ep}$ are
not entirely fixed by reparameterisation invariance\bib{rand}.  The most
general form contains the 2 arbitrary numbers $\ga_{\ep}$ and
$\ga_{0}$ and the symmetric traceless matrix\footnote{Later we will
argue that $\M_{ab}$ is in fact not arbitrary, but is independent of
the regularisation scheme.} $\M_{ab}$ which only contains terms with
two derivatives
\begin{equation}
\left. \pl_{z_{1}^{a}}\pl_{z_{2}^{b}} \De_{\ep}(z_{1},z_{2})
\right|_{z_{1}=z_{2}=z} = \ga_{\ep}\de_{ab}\ep^{-2}e^{2\si} +
\ha\ga_{0}\de_{ab}\Box\si + \M_{ab} + 
O(\ep^{2})  \ .
\label{eqn.gen.sec.der}
\end{equation}
On the superficial level this looks disastrous since the equations of
motion may depend on the regularisation scheme used through the
numbers $\ga_{\ep}$ and $\ga_{0}$ ($\M_{ab}$ drops out of the
calculation at this level of the string).  In fact, it is
clear that all 
regularisation dependence can be soaked-up by redefining the dilaton
and the tachyon
\begin{eqnarray}
\phi'(p) & = & \phi(p) + \ha(\ga_{0} - 1)\eta^{\mu\nu}h_{\mu\nu} \mand
\non \\
T'(p) & = & T(p) + \ga_{\ep}\eta^{\mu\nu}h_{\mu\nu} \ .
\label{eqn.re.tach.dil}
\end{eqnarray}
It will soon become obvious that the factor of
$\ha\eta^{\mu\nu}h_{\mu\nu}$ serves 
to covariantise the equations of motion.  It is instructive to realise
that these field redefinitions
can be implemented by adding the local, worldsheet reparameterisation
invariant, term
\begin{equation}
\frac{1}{8\pi}\int \eta^{\mu\nu}h_{\mu\nu}\Box
\left.\De_{\ep}(z,z')\right|_{z'=z} \ ,
\end{equation}
to the action.   Equivalently, the covariant measure of
Eq.~(\ref{eqn.cov.m.m}) can be used
\begin{equation}
\left[\d^{D}\xi\right]_{\mbox{\scriptsize{cov}}} = \left[\d^{D}\xi
\sqrt{\det\ha G_{\mu\nu}}\right] \ .
\label{eqn.cov.meas}
\end{equation}
This measure has been previously considered by Andreev, 
Metsaev and 
Tseytlin\bib{amt}.  Regulating with the
short-distance cutoff, leads to
\begin{equation}
[\d^{D}\xi]_{\mbox{\scriptsize{cov}}} =
[\d^{D}\xi]\exp\left(-\frac{1}{8\pi}\int 
\d^{2}z \log\det G_{\mu\nu} \left.\Box\De_{\ep}(z,z')\right|_{z' =
z}\right) \ . 
\label{eqn.cm.reg}
\end{equation}
By performing a weak-field expansion of this new term and employing
the relation Eq.~(\ref{eqn.leib})
\begin{equation}
\pl_{a}\left(\left.\pl_{b}\De_{\ep}(z,z')\right|_{z=z'}\right)
= \left.\pl_{a}\pl_{b}\De_{\ep}(z,z')\right|_{z=z'}
 + \left.\pl_{b}\pl'_{a}\De_{\ep}(z,z')\right|_{z=z'} \ ,
\end{equation}
all regularisation ambiguities disappear.  So, either by field
redefinitions\footnote{in which case the tachyon and dilaton in
Eq.~(\ref{gen.fcn.after}) must be replaced by the redefined quantities $T'$ and
$\Phi'$ given by Eq.~(\ref{eqn.re.tach.dil})}, or by using the
covariant measure, the generating 
functional can be cast into the form
\begin{eqnarray}
Z[J] & = & P[\si]
\left( \frac{1}{V} \det{}' \frac{\Box}{4\pi} \right)^{-\ha D}
\left\{ \de^{D}(p^{\mu})	
- \frac{1}{4\pi} \int \d^{2}z e^{ipX_{0}} e^{(-p^{2}+ ip\cd Q)\si}
|\ep|^{p^{2}} \right. 
\non \\
&& \spa
\times \left[\ep^{-2} e^{2\si} T(p) + \Box \si (\phi(p) + \ha
\eta^{\mu\nu}h_{\mu\nu}) 
 + \ha\left(
\pl_{a}X^{\mu}\pl_{b}X^{\nu} + 2 i p^{\mu}\pl_{a}\si \pl_{b}X^{\nu} \right)
 \ep^{ab}B_{\mu\nu}(p)
\right.
\non \\
&& \spa\spa
+ \ha \left.\left.\left(
\pl_{a}X^{\mu}\pl_{a}X^{\nu} + 2 i p^{\mu} \pl_{a}\si \pl_{a}X^{\nu} 
- p^{\mu}p^{\nu} (\pl_{a}\si)^{2} \right) h_{\mu\nu}(p)
\right] \right\} \ . 
\label{gen.fcn.after}
\end{eqnarray}

Renormalisation at the linear level is trivial
\begin{equation}
T_{R}(p) = |\ep|^{p^{2}-2}T(p) \mand
(h_{R}^{\mu\nu}(p),B_{R}^{\mu\nu}(p),\phi_{R}(p)) =
|\ep|^{p^{2}}(h^{\mu\nu}(p),B^{\mu\nu}(p),\phi(p))  \ .
\end{equation}
This corresponds to a minimal subtraction scheme and can be clearly
implemented by adding local counter-terms to the action.  For
notational simplicity the subscripts $R$ will be dropped in what
follows. 

Finally, the limit $\ep\rightarrow 0$ can be taken and the generating
functional can be varied with respect to $\si$ to yield
\begin{eqnarray}
0 = \frac{\de Z[J]}{\de \si} & = & - 
P[\si]
\left( \frac{1}{V} \det{}' \frac{\Box}{4\pi} \right)^{-\ha D}
\frac{1}{4\pi}  e^{ipX_{0}} e^{(-p^{2}+ ip\cd Q)\si}
 \left\{ 
e^{2\si} \left( 2-  p^{2} + ip\cd Q\right) T \right.
\non \\
&& \spa
- \left[\Box \si + \pl_{a}\si\pl_{a}X_{0}^{\la}ip^{\la} +
(\pl_{a}\si)^{2}\ha(-p^{2} + ip\cd Q)\right]
\non \\
&& \spa\spa\spa \times
 \left[ \mbox{$\frac{D-26}{3}$} + (\eta_{\mu\nu} + h_{\mu\nu})Q^{\mu}Q^{\nu} -
2p^{2}\phi + R + 2ip\cd Q(\phi + \ha h) - 2ip^{\mu}Q^{\nu}h_{\mu\nu}
\right]
\non \\
&& \spa
+ \left( \ha \ep^{ab}\pl_{a}X^{\mu}\pl_{b}X^{\nu} +
\ep^{ab}\pl_{a}\si\pl_{b}X^{\mu}p^{\nu} \right)  3(-p^{\la} +
iQ^{\la})H_{\la\mu\nu} 
\non \\
&& \spa
+  \pl_{a}X_{0}^{\mu}\pl_{a}X_{0}^{\nu} \left(
 R_{\mu\nu} -  p_{\mu}p_{\nu}\Phi + \ha i p\cd Q h_{\mu\nu} -
ip_{\mu}Q^{\la}h_{\la\nu} \right)
\non \\
&& \spa\spa \left.
+    \Box X_{0}^{\mu} 
\left( ip^{\mu} (\phi + \ha h) -  (ip^{\nu} + Q^{\nu})h_{\mu\nu}
\right) \right\}\ ,
\label{eqn.set.to.z.m}
\end{eqnarray}
where $h = \eta^{\mu\nu}h_{\mu\nu}$ and
\begin{eqnarray}
2R_{\mu\nu} & = & - p^{2} h_{\mu\nu} - p_{\mu}p_{\nu} h +
p^{\la}p_{\mu}h_{\nu\la} + p^{\la}p_{\nu}h_{\mu\la}
\ , \non \\
H_{\la\mu\nu} & = & \thi(p_{\la}B_{\mu\nu} +
p_{\mu}B_{\nu\la} + p_{\nu}B_{\la\mu}) \ .
\end{eqnarray}
Identifying the coefficients of each linearly-independent term with
zero yields the standard linearised field equations (now expressed in
position coordinates) 
\begin{eqnarray}
0 & = & (\nabla^{2} + Q\cd \nabla + 2)T
 \ ,\non \\
0 & = & \mbox{$\frac{D-26}{3}$} + (\nabla_{\mu}\Phi)^{2} + 2 \nabla^{2}\Phi + R
\ ,\non \\
0 & = & (\nabla^{\la} + Q^{\la})H_{\la\mu\nu}
\ , \non \\
0 & = &  R_{\mu\nu} + \nabla_{\mu}\nabla_{\nu}\Phi \ ,
\label{eqn.mot.lin.1}
\end{eqnarray}
and the gauge condition
\begin{equation} 
0 = \pl_{\mu}(\phi + \ha h) - (\pl^{\nu} + Q^{\nu})h_{\mu\nu} \ .
\end{equation}
These equations are correct to first order in $\la$ and $\nabla_{\mu}$
is the covariant spacetime derivative.

A word can now be said about the connection of this work to the
standard approach in which the beta-functions are calculated and set
to zero.  In Eq.~(\ref{eqn.set.to.z.m}) the
beta functions are the coefficients of the linearly independent terms
$e^{2\si}$, $\Box\si$, $\ep^{ab}\pl_{a}X_{0}^{\mu}\pl_{b}X_{0}^{\nu}$
and $\pl_{a}X_{0}^{\mu}\pl_{a}X_{0}^{\nu}$.  However, there are also
`beta-functions' corresponding to terms that would have been non-local
in the original action, $(\pl\si)^{2}$ and
$\pl_{a}\si\pl_{b}X_{0}^{\mu}$.  It is not just luck that setting the
standard beta-functions to zero implies that these new `non-local
beta-functions' are zero too.  This can be verified by writing the
most general generating functional with two derivatives and varying it
with respect to $\si$.  The `beta-functions' corresponding to
$(\pl\si)^{2}$ and $\pl_{a}\si\pl_{b}X^{\mu}$ are always derivatives 
of the dilaton and antisymmetric-tensor beta-functions.  At the first
massive level this 
general argument no longer holds.  There are operators in $\de
Z/\de\si$, which correspond to non-local terms in the original action,
whose coefficients are not-necessarily derivatives of other
beta-functions.  However, although the general argument breaks down,
in practise, setting the beta-functions and the gauge constraints
equal to zero guarantees Weyl invariance of the theory.

\subsect{Higher-order corrections to the tachyon field equation}
In the final part of this section, the $T^{2}$ corrections to the
generating functional of Eq.~(\ref{gen.fcn.after}) will be discussed.
The path integral is easily evaluated to give a term proportional to
\begin{equation}
\int \d p_{1} T(p_{1})T(p_{2}) \int \d^{2}z_{1}\d^{2}z_{2}
f_{1}(z_{1})f_{2}(z_{2}) \exp \left( - p_{1}\cd p_{2}
\De_{\ep}(z_{1},z_{2}) \right) \ 
,
\end{equation}
in which $f_{i}(z_{i}) = \exp\left[ i p_{i}X(z_{i}) +
(2-p_{i}^{2})(\si(z_{i}) + \log|\ep|)\right]$ and $p_{1}^{\mu} + p_{2}^{\mu} +
J_{0}^{\mu}\sqrt{V} - 2Q^{\mu} = 0$.  Up to derivatives on
$\si$, the propagator can be written as
\begin{equation}
\De_{\ep}(z_{-},z_{+}) = -\log \left( 4|z_{-}|^{2} +
\ep^{2}e^{-2\si(z_{+})} \right) \ , 
\end{equation}
where $z_{\pm} = \ha(z_{1} \pm z_{2})$.  Expanding the $f_{i}$ around
$z_{-} = 0$, the integral over $z_{-}$ can now be performed.  Of
course this expansion is only valid for small $z_{-}$.  It is assumed
that the integral is finite because the worldsheet is compact and
that the $f_{i}$ are well behaved.  The integral is then dominated by
small $z_{-}$.

An aside can now be made explaining why the simple-minded method
employed in this paper is not suited to finding all the field
equations to quadratic order.  The expansion of the $f_{i}$ and higher
derivative terms in the 
regulated propagator generate terms with
arbitrary powers of $\pl_{a}\si$ and $\pl_{a} X^{\mu}$.  This means
that, generically, $T^{2}$ terms will appear in every field
equation.  This is true for every other field too:  The equation of
motion and linear constraints $C_{i}$ for an arbitrary field $F$ look
like 
\begin{equation}
(\nabla^{2} - M_{F}^{2})F = O(\la^{2}) \mand C_{i}(F) =
O(\la^{2}) \ ,
\end{equation}
to $O(\la^{2})$.  It would thus be impossible to calculate the
infinite number of $O(\la^{2})$ corrections.  Instead,
with a finite number of fields $F$ being $O(\la)$ and the rest
$\bar{F}$ being $O(\la^{2})$, it is {\em assumed} that the equations 
\begin{equation}
(\nabla^{2} - M_{\bar{F}}^{2})\bar{F} = O(F^{2}) \mand C_{i}(\bar{F})
= O(F^{2}) \ ,
\end{equation}
have a solution for $\bar{F}$.  Then, to $O(\la^{2})$, the only
equations that need 
be considered are
\begin{equation}
(\nabla^{2} - M_{F}^{2})F = O(F^{2}) \mand C_{i}(F) = O(F^{2}) \ .
\end{equation}

The $T^{2}$ contribution to the tachyon field equation is the
easiest to calculate and will be of use later when 1-1 tachyon
scattering is discussed.  The contribution has no derivatives on
$X^{\mu}$ and $\si$ and reads
\begin{equation}
Z_{TT} =   - P[\si]
\left(\frac{1}{V} \det{}' \frac{\Box}{4\pi} \right)^{-\ha D}
\int d^{2} z e^{ipX} e^{(2- p^{2})(\si - \log |\ep|)}
 \frac{1}{32\pi} \int \d p_{1}
\frac{T(p_{1})T(p - p_{1})}{1 +  p_{1}\cd(p-p_{1})} \ ,
\label{eqn.z.tt}
\end{equation}
where $X^{\mu}$ is still given by Eq.~(\ref{eqn.x.}) and $p^{\mu} +
\Im J_{0}^{\mu}\sqrt{V} = 0$.
The renormalisation condition must be modified to read
\begin{equation}
T_{R}(p) = |\ep|^{p^{2} -2}T(p) + \frac{1}{8}
\int \d p_{1} \frac{T(p_{1})T(p-p_{1})}{1 + p_{1}\cd (p-p_{1})}
\left(1 - |\ep|^{-2 -2p_{1}(p-p_{1})} \right) \ .
\end{equation}
Again, this can be implemented by adding local counterterms to the
action.  The
linearised equation of motion for the tachyon, $T(p)$, implies that
$-p^{2} + iQ\cd p + 2= 0$ to first order.  Thus, to this order, it is
correct to set 
$-p_{1}^{2} + iQ\cd p_{1} + 2 = 0 = -(p-p_{1})^{2} + iQ\cd (p-p_{1}) +
2$ in denominator of Eq.~(\ref{eqn.z.tt}).  Weyl invariance then gives
\begin{equation}
(\nabla^{2} + Q\cd\pl + 2)T - \qu T^{2} = 0 \ .
\end{equation}

\sect{The Massive Fields}
\label{sec.mass}

The analogous steps that were carried out at the massless level are now
performed at the first massive level of the string.
The most general action consistent with reparameterisation invariance,
both on the worldsheet and in spacetime, is gauge-fixed by eliminating
all St\"{u}ckelberg degrees of freedom.  Field 
redefinitions are employed in the path-integral to simplify the action
further and to 
eliminate regularisation ambiguities.  Renormalisation is performed
and the linearised equations of motion are calculated.   These are solved
in two spacetime dimensions.  A non-linear corrections to the tachyon
field equation is then derived.

\subsect{The linearised field equations for the first massive level}
It is well known that the first massive level of the closed bosonic
string consists 
of a field $E_{\mu\nu\la\rho} = E_{(\mu\nu)(\la\rho)}$ upon which the
Virasoro constraints impose traceless inside pairs of indices and
transversality\bib{wein}.  These conditions have not yet been derived
using the Wilson renormalisation group method\bib{jjj}.  Recently,
Bardakci and  
Bernardo\bib{bar1} have presented
an elegant procedure for deriving the field equations and constraints
due to Weyl invariance in which they retain some degree of manifest
covariance.  Using their formalism, all the massive levels have been
considered\bib{bar2}, however, to date the calculations have only been
performed in a flat empty background with $D\neq 26$. 

The most general  
reparameterisation invariant action with four derivatives on a curved
worldsheet has many gauge symmetries.   A systematic study of these
symmetries has been made by Buchbinder et.\ al\bib{buch}.  Gauge
invariance in string theory has also been extensively investigated by
Evans et. al\bib{evans}.  For each
gauge symmetry there is an associated St\"{u}ckelberg field.  Fixing
each St\"{u}ckelberg field to zero fixes its corresponding gauge.
For instance, the action contains the three terms
\begin{eqnarray}
&& \int \sqrt{g} \left( W_{\mu\nu\la\rho}(\xi)
\pl_{a}\xi^{\mu}\pl_{b}\xi^{\nu}\pl_{c}\xi^{\la}\pl_{d}\xi^{\rho}
g^{ab}g^{cd} + 
 A_{\mu\nu\la}(\xi) D^{2} \xi^{\mu}
\pl_{a}\xi^{\nu}\pl_{b}\xi^{\la}g^{ab} \right.
\non \\
&& \spa\spa
\left. + S_{\mu\nu\la}(\xi)
D_{a}\pl_{b}\xi^{\mu}\pl_{c}\xi^{\nu}\pl_{d}\xi^{\la}
g^{ac}g^{bd} \right) \ .
\end{eqnarray}
It is evident that this is invariant under
\begin{eqnarray}
\de W_{\mu\nu\la\rho} & = &
 \ha \pl_{\left(\mu\right.}\La_{\left.\nu\right)(\la\rho)}  +
 \ha \pl_{\left(\la\right.}\La_{\left.\rho\right)(\mu\nu)} 
\ , \non \\
\de A_{\mu\nu\la} & = & \La_{\mu(\nu\la)}
\ , \non \\
\de S_{\mu\nu\la} & = & \La_{\nu(\la\mu)} + \La_{\la(\nu\mu)}
 \ ,
\end{eqnarray}
in which symmeterisation is indicated with round brackets, for example
$\La_{\mu(\nu\la)} = \ha\La_{\mu\nu\la} + \ha\La_{\mu\la\nu}$.  In
this case, the field $S_{\mu\nu\la}$ can be considered a
St\"{u}ckelberg field and eliminated by choosing an appropriate
$\La_{\mu\nu\la}$.  
After a complete gauge-fixing, the remaining
terms can be written 
\begin{eqnarray}
S_{M}(\xi^{\mu},g_{ab}) & =  &
\int \sqrt{g} W_{\mu\nu\la\rho}(\xi)
\pl_{a}\xi^{\mu}\pl_{b}\xi^{\nu}\pl_{c}\xi^{\la}\pl_{d}\xi^{\rho}
g^{ab}g^{cd} +
\sqrt{g}RW_{\mu\nu}(\xi)\pl_{a}\xi^{\mu}\pl_{b}\xi^{\nu}
g^{ab} + \sqrt{g}R^{2}W(\xi)
\non \\
&& \spa
+  \int \bar{W}_{\mu\nu\la\rho}(\xi)
\pl_{a}\xi^{\mu}\pl_{b}\xi^{\nu}\pl_{c}\xi^{\la}\pl_{d}\xi^{\rho}
g^{ab}i\ep^{cd} +
R\bar{W}_{\mu\nu}(\xi)\pl_{a}\xi^{\mu}\pl_{b}\xi^{\nu}i\ep^{ab} \ .
\non \\
&& \spa
+  \int  \sqrt{g}RD^{2} \xi^{\mu}A_{\mu}(\xi)
+ \sqrt{g}D^{2}\xi^{\mu}D^{2}\xi^{\nu}A_{\mu\nu}(\xi)
+ \sqrt{g} A_{\mu\nu\la}(\xi) D^{2} \xi^{\mu}
\pl_{a}\xi^{\nu}\pl_{b}\xi^{\la}g^{ab}
\non \\
&& \spa
+ \int \bar{A}_{\mu\nu\la}(\xi) D^{2}\xi^{\mu}
\pl_{a}\xi^{\nu}\pl_{b}\xi^{\la}i\ep^{ab}
 \ .
\end{eqnarray}

As in the massless case, the $A$-type fields can be shifted away by a
change of variables in the path integral.  Specifically, the analogue
of the shift in Eq.~(\ref{eqn.shift.tildA}) is
\begin{equation}
\xi'{}^{\mu} = \xi^{\mu} - RA^{\mu} - A^{\mu}{}_{\nu}D^{2}\xi^{\nu} -
RA^{\mu}{}_{\nu}Q^{\nu} -
A^{\mu}{}_{\nu\la}\pl_{a}\xi^{\nu}\pl_{b}\xi^{\la}g^{ab}
-
\bar{A}^{\mu}{}_{\nu\la}\pl_{a}\xi^{\nu}\pl_{b}\xi^{\la}i\ep^{ab}/\sqrt{g}
\ ,
\end{equation}
with covariant measure being
\begin{eqnarray}
\left[\d^{D}\xi\right]_{\mbox{\scriptsize{cov}}} & = &
\left[\d^{D}\xi \det{}^{1/2}   \left(\ha G_{\mu\nu} -
R\nabla_{\left(\mu\right.}A_{\left.\nu\right)} - 
\nabla_{\left(\mu\right.}(A_{\left.\nu\right)\la}D^{2}\xi^{\la}) 
- R\nabla_{\left(\mu\right.}(A_{\left.\nu\right)\la}Q^{\la}) 
\right. \right.
\non \\
&&\spa\spa
\left.\left.
- \nabla_{\left(\mu\right.}(A_{\left.\nu\right)\la\rho}
\pl_{a}\xi^{\la}\pl_{b}\xi^{\rho}) 
- \nabla_{\left(\mu\right.}(\bar{A}_{\left.\nu\right)\la\rho}
\pl_{a}\xi^{\la}\pl_{b}\xi^{\rho} 
i\ep^{ab}/\sqrt{g}) \right)\right] \ .
\end{eqnarray}
The $W$-type fields also need to be redefined just as the dilaton
$\phi$ was shifted to $\phi + Q\cd \tilde{A}$ in
Eq.~(\ref{eqn.shift.tildA}). 

The remaining terms in the action can be grouped together in a more
compact form by using the 2-dimensional identity
\begin{equation}
\ep^{ab}\ep^{cd}g_{ac}g_{bd} = 2g \ .
\end{equation}
Defining $f^{ab} = \sqrt{g}g^{ab} + i\ep^{ab}$ this identity leads to
the following symmetries of a product of two $f^{ab}$ densities
\begin{equation}
f^{ac}f^{bd} = f^{bc}f^{ad} 
\mand f^{ac}f^{bd}g_{ab} = 0 \ .
\label{eqn.f.sym.tr}
\end{equation}
Then the action at the first massive level can be written in the
standard fashion 
\begin{equation}
S_{M}(\xi^{\mu},g_{ab}) =
\int {g}^{-1/2} \left (E_{\mu\nu\la\rho}(\xi)
\pl_{a}\xi^{\mu}\pl_{b}\xi^{\nu}\pl_{c}\xi^{\la}\pl_{d}\xi^{\rho}
f^{ac}f^{bd} + 
\sqrt{g}RE_{\mu\nu}(\xi)\pl_{a}\xi^{\mu}\pl_{b}\xi^{\nu}
f^{ab} + (\sqrt{g}R)^{2}E(\xi) \right) \ ,
\end{equation}
in which
\begin{eqnarray}
2E_{\mu\nu\la\rho} & \equiv &  W_{\mu\la\nu\rho} +
W_{\mu\rho\nu\la} + \bar{W}_{\mu\la\nu\rho} + \bar{W}_{\nu\rho\mu\la}
\ , \non \\
E_{\mu\nu} & \equiv & W_{\mu\nu} + \bar{W}_{\mu\nu}
\ , \non \\
E & \equiv & W \ ,
\end{eqnarray}
The field $E_{\mu\nu\la\rho}$ is symmetric in pairs of indices,
$E_{\mu\nu\la\rho} = E_{(\mu\nu)(\la\rho)}$, which can be seen by using
Eq.~(\ref{eqn.f.sym.tr}).

The partition function at the
linear level is
\begin{eqnarray}
Z[J] & = &  P[\si]
\left( \frac{1}{V} \det{}' \frac{\Box}{4\pi} \right)^{-\ha D}
\left\{ \de^{D}(p^{\mu})	
- \frac{1}{4\pi} \int \d^{2}z e^{ipX} g^{-1/2}e^{- \ha
p^{2}\De_{\ep}(z,z)}\ep^{2} \right.
\non \\
&& 
\times \left[
E_{\mu\nu\la\rho}f^{ac}f^{bd} \left(
\pl_{a}X^{\mu}\pl_{b}X^{\nu}\pl_{c}X^{\la}\pl_{d}X^{\rho}
+ 2ip^{\mu}\pl_{a}\De\pl_{b}X^{\nu}\pl_{c}X^{\la}\pl_{d}X^{\rho}
+ 2ip^{\la}\pl_{a}X^{\mu}\pl_{b}X^{\nu}\pl_{c}\De\pl_{d}X^{\rho}
\right. \right.
\non \\
&& \spa
+(\eta^{\mu\nu}\pl_{a}\pl'_{b}\De - p^{\mu}p^{\nu}\pl_{a}\De\pl_{b}\De)
\pl_{c}X^{\la}\pl_{d}X^{\rho}
+(\eta^{\la\rho}\pl_{c}\pl'_{d}\De - p^{\la}p^{\rho}\pl_{c}\De\pl_{d}\De)
\pl_{a}X^{\mu}\pl_{b}X^{\nu}
\non \\
&& \spa
+ 4(\eta^{\mu\la}\pl_{a}\pl'_{c}\De - p^{\mu}p^{\la}\pl_{a}\De\pl_{c}\De)
\pl_{b}X^{\nu}\pl_{d}X^{\rho}
\non \\
&& \spa
+ 2\pl_{a}X^{\mu}(2\eta^{\nu\la}ip^{\rho}\pl_{b}\pl'_{c}\De\pl_{d}\De
+ \eta^{\la\rho}ip^{\nu}\pl_{c}\pl'_{d}\De\pl_{b}\De -
ip^{\nu}p^{\la}p^{\rho}\pl_{b}\De\pl_{c}\De\pl_{d}\De) 
\non \\
&& \spa
+ 2(\eta^{\mu\nu}ip^{\la}\pl_{a}\pl'_{b}\De\pl_{c}\De +
2\eta^{\mu\la}ip^{\nu}\pl_{a}\pl'_{c}\De\pl_{b}\De -
ip^{\mu}p^{\nu}p^{\la}\pl_{a}\De\pl_{b}\De\pl_{c}\De) \pl_{d}X^{\rho}
\non \\
&& \spa
+ \eta^{\mu\nu}\eta^{\la\rho} \pl_{a}\pl'_{b}\De\pl_{c}\pl'_{d}\De +
2\eta^{\mu\la}\eta^{\nu\rho}\pl_{a}\pl'_{c}\De\pl_{b}\pl'_{d}\De 
\non \\
&& \spa
- \eta^{\mu\nu}p^{\la}p^{\rho}\pl_{a}\pl'_{b}\De\pl_{c}\De\pl_{d}\De 
- 4\eta^{\mu\la}p^{\nu}p^{\rho}\pl_{a}\pl'_{c}\De\pl_{b}\De\pl_{b}\De
- \eta^{\la\rho}p^{\mu}p^{\nu}\pl_{c}\pl'_{d}\De\pl_{a}\De\pl_{b}\De 
\non \\
&& \spa
\left.
+ p^{\mu}p^{\nu}p^{\la}p^{\rho}\pl_{a}\De\pl_{b}\De\pl_{c}\De\pl_{d}\De
\right)
\non \\
&& \spa
+ \sqrt{g}RE_{\mu\nu}f^{ab} \left(
\pl_{a}X^{\mu}\pl_{b}X^{\nu} + ip^{\mu}\pl_{a}\De\pl_{b}X^{\nu} +
ip^{\nu}\pl_{a}X^{\mu}\pl_{b}\De + \eta^{\mu\nu}\pl_{a}\pl'_{b}\De -
p^{\mu}p^{\nu}\pl_{a}\De\pl_{b}\De \right)
\non \\
&& \spa
+ \left. (\sqrt{g}R)^{2} \left. E \right] \right\} \ .
\label{eqn.gf.e}
\end{eqnarray}
In this equation, the spacetime fields have been scaled by $\ep^{2}$
and the shorthand notation $\pl_{a}\De$ and 
$\pl_{a}\pl'_{b}\De$ has been used for the expressions
$\left.\pl_{z^{a}}\De(z,z')\right|_{z'=z}$ and
$\left.\pl_{z^{a}}\pl_{z'{}^{b}}\De(z,z')\right|_{z'=z}$
respectively.

Once again it looks as if regularisation ambiguities
may be a problem, because the second derivative of the regularised
propagator, given by Eq.~(\ref{eqn.gen.sec.der}), contains the
arbitrary numbers $\ga_{\ep}$ and $\ga_{0}$ and the symmetric
traceless matrix $\M_{ab}$.   The situation is
complicated further by the terms that look generically like $\pl_{a}\pl_{b}'\De
\pl_{c}\De$ and $\pl_{b}\pl_{b}'\De\pl_{c}\pl_{d}'\De$.  Recall that
$\pl_{a}\pl_{b}'\De$ is of order 
$\ep^{-2}$ so that in 
such terms, $O(\ep^{2})$ corrections to $\pl_{a}\De$ and
$\pl_{a}\pl_{b}'\De$ must 
be considered.   In fact, most ambiguities can be absorbed by
following the same line of thought that lead to the 
covariant measure of Eq.~(\ref{eqn.cov.meas}).  Consider first the terms
\begin{eqnarray}
&&\left[
E_{\mu\nu\la\rho}f^{ac}f^{bd}
4\eta^{\mu\la}\pl_{a}\pl_{c}'\De \left(
\pl_{b}X^{\nu}\pl_{d}X^{\rho}
+ ip^{\nu}\pl_{b}\De\pl_{d}X^{\rho}
+ ip^{\rho}\pl_{b}X^{\nu}\pl_{d}\De
\right.\right.
\non \\
&& \spa\spa\spa
\left.\left.
+ \ha \eta^{\nu\rho}\pl_{b}\pl_{d}'\De
- p^{\nu}p^{\rho}\pl_{b}\De\pl_{d}\De \right)
+ \sqrt{g}RE_{\mu\nu}f^{ab}\eta^{\mu\nu}\pl_{a}\pl_{b}'\De 
\right] \ .
\label{eqn.first.amb}
\end{eqnarray}
By using the covariant measure 
\begin{eqnarray}
\left[\d^{D}\xi\right]_{\mbox{\scriptsize{cov}}} & = &
\left[\d^{D}\xi \sqrt{
\det \left(\ha G_{\mu\nu} + RE_{\mu\nu} +
4E_{\mu\la\nu\rho}f^{ab}
(\pl_{a}X^{\la}\pl_{b}X^{\rho} - \ha
G^{\la\rho}\pl_{a}\pl_{b}\De_{\ep})/\sqrt{g}
\right)}\right] \ ,
\end{eqnarray}
and employing the Leibnitz-like relation Eq.~(\ref{eqn.leib}), the
ambiguities are removed from Eq.~(\ref{eqn.first.amb}).  Upon regulating
this measure as in Eq.~(\ref{eqn.cm.reg}), it is seen that the
extra terms added to the action are both local and worldsheet
reparameterisation invariant (and thus correspond to simple field
redefinitions as in 
Eq.~(\ref{eqn.re.tach.dil})).

However, there are more ambiguous
terms in the partition function.  These are
\begin{eqnarray}
&&\left[
E_{\mu\nu\la\rho} f^{ac}f^{bd}
\left\{
\eta^{\mu\nu} \pl_{a}\pl_{b}'\De
\left(\pl_{c}X^{\la}\pl_{d}X^{\rho}
+ ip^{\la}\pl_{c}\De\pl_{d}X^{\rho}
+ ip^{\rho}\pl_{c}X^{\la}\pl_{d}\De
\right.\right. \right.
\non \\
&& \spa\spa\spa
\left.\left. \left.
+ \ha \eta^{\la\rho}\pl_{c}\pl_{d}'\De
- p^{\la}p^{\rho}\pl_{c}\De\pl_{d}\De \right)
+ (\mu\nu ab) \leftrightarrow (\la\rho cd) \right\}
\right] \ .
\label{eqn.second.amb}
\end{eqnarray}
Since $f^{ac}f^{bd}$ is symmetric and traceless in $ab$ (see
Eq.~(\ref{eqn.f.sym.tr})), the $\pl_{a}\pl_{b}'\De$ becomes simply
$\M_{ab}$.  To remove this ambiguity, the
term
\begin{equation}
gE_{\mu\nu\la\rho} f^{ac}f^{bd} \left\{
\eta^{\mu\nu}
\M_{ab}
\pl_{c}X^{\la}\pl_{d}X^{\rho}
+ (\mu\nu ab) \leftrightarrow (\la\rho cd) \right\}
 \ ,
\end{equation}
would have to be added to the Lagrangian.  Unless $\M_{ab}$ is zero
then, this term is not a local counterterm since it will contain
$\pl_{a}\pl_{b}\si$ or $\pl_{a}\si\pl_{b}\si$.  However, one
particular diffeomorphism 
covariant calculation\bib{dhp} yields a non-zero value for $\M_{ab}$
\begin{equation}
\M_{ab} = \thi\left(\pl_{a}\pl_{b}\si + \pl_{a}\si\pl_{b}\si - \ha
\de_{ab}\Box\si - \ha\de_{ab}(\pl\si)^{2} \right) \ .
\label{eqn.sym.tr}
\end{equation}
Since a local counterterm to remove $\M_{ab}$ cannot be found, we
propose that its non-vanishing is {\em independent} 
of the regularisation scheme used.

It is worth emphasising this point given the discussion in
Refs~\ref{jjj} and~\ref{buch} where the traceless condition on
$E_{\mu\nu\la\rho}$ was missed.  It is now clear that tracelessness
would be lost if there was a
reparameterisation invariant regularisation scheme in which
$\M_{ab} = 0$, since
Eq.~(\ref{eqn.second.amb}) would vanish and there are no other terms
in the partition function that depend on $\eta^{\mu\nu}$ and
$\eta^{\la\rho}$.   Notice that this can be seen in a
vertex-operator\bib{wein} calculation 
too --- tracelessness would come from a self-contraction of $\pl X^{\mu}\pl
X^{\nu}$, and if the regularisation was such that this was zero, no
tracelessness condition would be found.

After performing minimal subtraction
\begin{equation}
\left( E_{R}^{\mu\nu\la\rho},E_{R}^{\mu\nu},E_{R} \right) =
|\ep|^{p^{2} + 2}
\left( E^{\mu\nu\la\rho},E^{\mu\nu},E \right)  \ ,
\end{equation}
and taking the limit $\ep \rightarrow 0$, Weyl invariance can be
imposed on the partition function and the equations of motion obtained
\begin{eqnarray}
0 & = & (\nabla^{2} + Q\cd\nabla - 2)E_{\mu\nu\la\rho}
\ , \non \\
0 & = & \eta^{\mu\nu}E_{\mu\nu\la\rho} =
\eta^{\la\rho}E_{\mu\nu\la\rho}
\ , \non \\
0 & = & (\nabla^{\mu} + Q^{\mu})E_{\mu\nu\la\rho} = (\nabla^{\la} +
Q^{\la})E_{\mu\nu\la\rho} 
\ , \non \\
0 & = & \eta^{\mu\la}E_{\mu\nu\la\rho} + E_{\nu\rho}
\ , \non \\
0 & = & \eta^{\mu\nu}E_{\mu\nu} + 4 E \ .
\label{eqn.mot.e}
\end{eqnarray}
Evidently $E'_{\mu\nu}$ and $E'$
are just traces of $E_{\mu\nu\la\rho}$.  $E_{\mu\nu\la\rho}$ is
transverse and traceless inside the pairs of 
indices as expected.

In two spacetime dimensions, these equations can be solved to find a
higher-mode version of the black-hole.  The metric $\eta^{\mu\nu} =
\diag(\eta^{TT},\eta^{XX}) = \diag(-1,1)$ is used.  The most general
solution of equations of 
motion is given by
\begin{equation}
E_{\mu\nu\la\rho} = 
\left\{
\begin{array}{ll}
E_{+} & \mbox{if $\mu + \nu + \la + \rho = \mbox{even}$}
\\
E_{-} & \mbox{if $\mu + \nu + \la + \rho = \mbox{odd}$}
\end{array}
\right. \ ,
\end{equation}
where
\begin{equation}
E_{\pm} = Ae^{-\left(Q^{X} + \frac{2}{Q^{X} + Q^{T}}\right)X +
\left(Q^{T} - 
\frac{2}{Q^{X} + Q^{T}}\right)T} \pm B
e^{-\left(Q^{X} + \frac{2}{Q^{X} - Q^{T}}\right)X + \left(Q^{T} +
\frac{2}{Q^{X} - Q^{T}}\right)T} \ , 
\label{eqn.background.e}
\end{equation}
in which $A$ and $B$ are arbitrary constants of integration.  In the
non-generic case  
of $Q^{X} \pm Q^{T} = 0$ the term which contains the potential
infinity must be dropped from the solution.  Thus, for generic
background charge, the background solution of $E_{\mu\nu\la\rho}$ is a
time-dependent, two-parameter solution.

\subsect{Massive-field corrections to the tachyon field equation}
In a flat linear-dilaton background $0 = h'_{\mu\nu} = B'_{\mu\nu} =
\phi'$, the linearised equations of motion simply consist of
the equations for the massive field $E_{\mu\nu\la\rho}$ given
by Eq.~(\ref{eqn.mot.e}), the dilaton field equation $Q^{2} =
(26-D)/3$ and the tachyon equation.  This section considers
higher-order corrections to the tachyon field equation in this
background.   It is sufficient to calculate the $T^{2}$ and $TE$
contributions in order to make a comparison with
the proposal of DMW.  The
$T^{2}$ part has already been calculated in Sec.~\ref{sec.method}.
There are two ways in which a $TE$ term can be obtained.  One comes
from the covariant measure
\begin{eqnarray}
&&   \frac{1}{16\pi^{2}} \int \left[\d^{D}\xi\right]
\exp \left( -\mbox{$\frac{1}{8\pi}$} \int
\pl_{a}\xi^{\mu}\pl_{a}\xi^{\nu}\eta_{\mu\nu} +
\mbox{$\frac{1}{4\pi}$}a_{0}\cd (J_{0} - \mbox{$\frac{2Q}{\sqrt{V}}$}) \right)
\int \d^{2}z_{1}\d^{2}z_{2}\d^{D}p_{1}\d^{D}p_{2}
\non \\
&& \spa\spa\spa\spa
\times
 e^{i(p_{1}(\xi+ X)(z_{1}) +
p_{2}(\xi + X)(z_{2}))}
e^{2\si(z_{1})}T(p_{1})
\non \\
&& \spa\spa\spa\spa
\times
 4
\eta^{\mu\la}E_{\mu\nu\la\rho}(p_{2})f^{ab}/\sqrt{g} \left(
\pl_{a}\xi^{\nu}(z_{2})\pl_{b}\xi^{\rho}(z_{2})
- \ha\eta^{\nu\rho}\pl_{a}\pl_{b}\De \right)(-2\ga_{\ep}\ep^{-2}e^{2\si})  \ ,
\label{eqn.te.redef}
\end{eqnarray}
while the other is simply
\begin{eqnarray}
&&  \frac{1}{16\pi^{2}} \int \left[\d^{D}\xi\right]
\exp \left( -\mbox{$\frac{1}{8\pi}$} \int
\pl_{a}\xi^{\mu}\pl_{a}\xi^{\nu}\eta_{\mu\nu} +
\mbox{$\frac{1}{4\pi}$}a_{0}\cd (J_{0} - \mbox{$\frac{2Q}{\sqrt{V}}$})
\right)\int \d^{2}z_{1}\d^{2}z_{2}\d^{D}p_{1}\d^{D}p_{2}
\non \\
&& \spa\spa\spa\spa
\times
e^{i(p_{1}(\xi + X)(z_{1}) + 
p_{2}(\xi + X)(z_{2}))}
e^{2\si(z_{1})-2\si(z_{2})}
T(p_{1})
\non \\
&& \spa\spa\spa\spa
\times
E_{\mu\nu\la\rho}(p_{2}) f^{ac}f^{bd}
\pl_{a}\xi^{\mu}(z_{2})\pl_{b}\xi^{\nu}(z_{2})
\pl_{c}\xi^{\la}(z_{2})\pl_{d}\xi^{\rho}(z_{2})  \ .
\label{eqn.te.real}
\end{eqnarray}
All other terms give derivatives on $\si$ or $X^{\mu}$ which will
contribute to other field equations, as explained in
Sec.~\ref{sec.method}.  The result 
of performing these Gaussian integrals is quite simple because all
terms containing derivatives on $\si$ or $X^{\mu}$ can be dropped.
By construction, the parts from the covariant measure exactly cancel
the regularisation dependent parts in Eq.~(\ref{eqn.te.real}), leading
to a total 
contribution of
\begin{eqnarray}
&&\frac{1}{16\pi^{2}} \int \d^{2}z_{1}\d^{2}z_{2}\d^{D}p_{1}
e^{i(p_{1}X_{1} + p_{2}X_{2})}e^{2\si_{1} -
2\si_{2}}T(p_{1})E_{\mu\nu\la\rho}(p_{2})
\non \\
&& \spa\spa\spa\spa\spa
p_{1}^{\mu}p_{1}^{\nu}p_{1}^{\la}p_{1}^{\rho}
 (\pl_{z_{2}^{a}}\De(z_{1},z_{2}))^{4} 
e^{-\ha p_{1}^{2} \De(z_{1},z_{1}) - \ha
p_{2}^{2}\De(z_{2},z_{2}) -  p_{1}\cd p_{2}\De(z_{1},z_{2}) }\ ,
\end{eqnarray}
with momentum conservation $p_{1} + p_{2} + J_{0}^{\mu}\sqrt{V} - 2Q^{\mu} =
p_{1} + p_{2} + p = 0$.  In Sec.~\ref{sec.method}, the $TT$ term was
simplified by 
performing the 
integral over $z_{-}$.  This integral can also be done here.  Using
\begin{equation}
(4\pi\pl'_{a}\De(z,z'))^{2} = \frac{4|z-z'|^{2}}{(|z-z'|^{2} +
\ep^{2}e^{-2\si})^{2}} \ ,
\end{equation}
which is true up to derivatives on $\si$, the $TE$ term finally reads
\begin{eqnarray}
Z_{TE} & = & P[\si]
\left(\frac{1}{V} \det{}' \frac{\Box}{4\pi} \right)^{-\ha D}
\int d^{2} z e^{ipX} e^{(2- p^{2})(\si - \log |\ep|)}
\non \\
&& \spa\spa\spa
 \times \frac{2}{\pi} \int \d p_{1}
\frac{E_{\mu\nu\la\rho}(p-p_{1})
  p_{1}^{\mu}p_{1}^{\nu}p_{1}^{\la}p_{1}^{\rho}T(p_{1})}
  {(3-p_{1}(p-p_{1}))(2-p_{1}(p-p_{1}))(1-p_{1}(p-p_{1}))}\ .
\label{eqn.z.te}
\end{eqnarray}

After renormalisation, the denominators of
Eqs.~(\ref{eqn.z.tt})~and~(\ref{eqn.z.te}) can be simplified by
substituting  in the mass-shell relations
\begin{equation}
(-p^{2} + iQp + 2)T(p) + O(\la^{2}) = 0 = (-p^{2} + iQp
-2)E_{\mu\nu\la\rho}(p)  + O(\la^{2})  \ .
\end{equation}
Imposing Weyl invariance and expanding the $TE$ contribution around
$2+ipQ - p^{2} = 0$, the field equation is finally obtained:
\begin{equation}
(\nabla^{2} + Q\cd\nabla + 2)T - \qu T^{2} +
8E^{\mu\nu\la\rho}\nabla_{\mu}\nabla_{\nu}\nabla_{\la}\nabla_{\rho}T = 0 \ .
\label{eqn.tach.e.theory}
\end{equation}
This equation is valid up to $O(E^{2})$ with a flat linear-dilaton
background.  All higher derivative terms in the expansion of the $TE$
term around $2+ipQ-p^{2}=0$ have been left out.  It will soon be
explained why these terms have no bearing on the validity of DMW's
proposal. 

\sect{Tachyon Scattering in the Matrix Model}
\label{sec.dmw}

As mentioned in the introduction, the double-scaling limit of the
matrix model defines the nonperturbative physics of 2D string theory.
After taking the limit, the matrix model consists of nonrelativistic,
non-interacting fermions living in an inverted harmonic potential.
The ground state consists of a flat sea filled up to the Fermi-surface
which lies $O(1/g_{\mbox{\scriptsize{str}}})$ below the top of the
potential.  The tachyons in spacetime have been found to be the
bosonised fluctuations of the Fermi surface.  To reproduce the
perturbative string theory results only very small bumps on top of the
Fermi sea need be considered.  Tunnelling through the potential barrier
and fluid washing over the wall are both non-perturbative effects.  So,
when working in the
semi-classical limit with very small pulses, it seems safe to work with
only one side of the potential.  However, DMW
argued that the spacetime metric must couple to the entire
energy-momentum tensor of the theory.  The Hamiltonian of the string
theory is identical to that of the matrix model (unless the potential
is modified by hand from the beginning).  A generic perturbation of the
Fermi fluid has total energy coming from both sides of the potential,
so the metric must couple to this total energy.  Allowing excitations
of both halves of the sea introduced another
degree of freedom into the model that had not yet been utilised.
There were now {\em two} scalar fields --- the average and the
difference of the bosonised fluctuations on each side of the barrier.
Comparing with the
tachyon-graviton effective theory, they found that the tachyon, $T$, is the
leg-pole transform of the average of the fluid fluctuations, while the
total energy of the difference variable, $\De$, could be identified with the
mass of the blackhole.

Specifically, after scaling $T = e^{-2x}S$,
the massless field $S$ satisfied the equation of motion (see DMW
Eq.~(5.4)) 
\begin{equation}
\pl_{+}\pl_{-}X = Me^{-4x}O_{x,t}S + e^{-2x}S_{0}S \ ,
\label{dmw.5.4}
\end{equation}
where $O_{x,t}$ is some second-order differential operator whose
details are not 
important, $S_{0}$ is the tachyon background and $x_{\pm} = t\pm x$.
 The first term originates from the $Th$ term while the
second comes from the $TT$ term.  The factor of $Me^{-4x}$ comes from
the blackhole solution to the graviton equation of motion, $h_{\mu\nu}
= Me^{-4x}$ in which $M$ is a constant and called the `mass' of the
blackhole.  Integrating the equation of motion
to first order in the 
background $S_{0}$ with the boundary condition $S(x,t)
\stackrel{t\rightarrow - \infty}{\longrightarrow}
S_{\mbox{\scriptsize{in}}}(x^{+})$ they obtained
\begin{eqnarray}
S_{\mbox{\scriptsize{out}}}(x^{-}) & = &
e^{2x^{-}}\int_{-\infty}^{\infty}\d x^{+}
e^{-x^{+}}\int_{-\infty}^{x^{-}} \d u^{-}
e^{u^{-}}S_{0}(u^{-},x^{+})S_{\mbox{\scriptsize{in}}}(x^{+})
\non \\
&& \spa\spa - \ha M e^{2x^{-}}\int_{-\infty}^{\infty}e^{-2x^{+}}
S_{\mbox{\scriptsize{in}}}(x^{+}) \ .
\label{eqn.sout}
\end{eqnarray}
The first term describes tachyon
scattering in the presence of the background $S_{0}$ and the second is
due to scattering off the blackhole background.  Mathematically the
second term comes about by going by parts with the operator $O_{x,t}$
under the integrals $\int_{-\infty}^{x^{-}}\d
u^{-}\int_{-\infty}^{\infty}\d x^{+}$.  The matrix model
predicts that the term describing tachyon scattering off a blackhole
background is proportional to
\begin{equation}
\int_{-\infty}^{\infty}\d \tau \,\De^{2}(\tau)\ 
e^{2x^{-}}\int_{-\infty}^{\infty}\d x^{+} e^{-2x^{+}}S_{0}(x^{+}) \ ,
\end{equation}
This lead DMW to
conclude that the mass of the blackhole was equal to the energy
contained in the difference variable $\int_{-\infty}^{\infty}\d \tau
\De^{2}(\tau)$.

Having found the effective theory describing
tachyon-$E_{\mu\nu\la\rho}$ scattering, the same steps can be followed
to find an equivalent to Eq.~(\ref{eqn.sout}).  Taking the dilaton to
lie purely 
in the $X^{1} = X$ direction means $Q^{X} = 2\sqrt{2}$.  DMW's conventions
(indicated by lower-case letters) differ slightly from those used in
this paper, the connection is $x^{\mu} = X^{\mu}/\sqrt{2}$.
The massless 
tachyon $S$ is related to $T$ by $T = e^{- Q\cd X/2}S = e^{-2x}S$.
Inserting this into the equation of motion for the tachyon yields the
equivalent of Eq.~(\ref{dmw.5.4})
\begin{equation}
\pl_{+}\pl_{-}S = O_{x,t}E_{\mu\nu\la\rho}O^{\mu\nu\la\rho}_{x,t}S +
\sqrt{2}e^{-2x}S_{0}S \ .
\end{equation}
The $\sqrt{2}$ in front of the $S_{0}S$ term is arbitrary since it can
be tuned by rescaling $T$, and the differential operators $O_{x,t}$ and
$O_{x,t}^{\mu\nu\la\rho}$ occur through expanding the denominators of
the $TE$ term in small $2+ipQ-p^{2}$.
Substituting the background solution for the $E$-field in
Eq.~(\ref{eqn.background.e}) 
and integrating this equation with the same boundary condition leads
to 
\begin{eqnarray}
S_{\mbox{\scriptsize{out}}}(x^{-}) & = &
e^{2x^{-}}\int_{-\infty}^{\infty}\d x^{+}
e^{-x^{+}}\int_{-\infty}^{x^{-}}
e^{u^{-}}S_{0}(u^{-},x^{+})S_{\mbox{\scriptsize{in}}}(x^{+})
\non \\
&& \spa\spa + c_{1}A e^{2x^{-}}\int_{-\infty}^{\infty}e^{-3x^{+}}
S_{\mbox{\scriptsize{in}}}(x^{+})
 + c_{2}Be^{3x^{-}}\int_{-\infty}^{\infty}e^{-2x^{+}}
S_{\mbox{\scriptsize{in}}}(x^{+})
 \ .
\label{eqn.tach.scat.e}
\end{eqnarray}
The numbers $c_{i}$ come from going by parts with the operator
$O^{\mu\nu\la\rho}_{x,t}$ and by applying $O_{x,t}$ to the background
solution for $E_{\mu\nu\la\rho}$.  DMW's representation of the matrix
model predicts that the terms describing tachyon scattering off a
discrete-mode background are proportional to
\begin{equation}
\int_{-\infty}^{\infty}\d \tau \De^{2}(\tau)\ e^{(m-n)\tau}
e^{(n+1)x^{-}} \int_{-\infty}^{\infty}\d x^{+} e^{-(m+1)x^{+}}
S_{\mbox{\scriptsize{in}}}(x^{+}) \ ,
\label{eqn.dmw.pred}
\end{equation}
where $m,n > 0$.
The $m=n=1$ term is the blackhole background.  By comparing this
equation with Eq.~(\ref{eqn.tach.scat.e}) it is
clear that the $(m,n) = (2,1)$ and $(m,n) = (1,2)$ can be identified
with the $A$ and $B$ terms of the first massive level of the string.
The $c_{i}$ (and thus the extra terms from the expansion in small
$2+ipQ - p^{2}$) just change the constants of proportionality in the
identification
\begin{equation}
A \propto \int_{-\infty}^{\infty}\d\tau \De^{2}(\tau) e^{\tau}
\mand
B \propto \int_{-\infty}^{\infty}\d\tau \De^{2}(\tau) e^{-\tau} \ .
\end{equation}
Thus, DMW's proposal has been checked to the first massive level in
the string spectrum.

\sect{Conclusion}
\label{sec.concl}

By imposing Weyl invariance on the generating functional of closed
bosonic string theory, the linearised equation of motion and
constraints for the first massive level, in a flat linear-dilaton
background, have been derived.  This is the first time that the
correct equations have been obtained using a variant of the Wilson
renormalisation group method.

In two spacetime dimensions, these constraints were solved
to find a two-parameter time-dependent solution.  One-to-one tachyon
scattering in this discrete-state background was studied and the 
results agree with the prediction made by Dhar, Mandal and
Wadia's\bib{dmw}$^{,}$\bib{dmw2} representation of
the matrix model.

It would be a thankless task to derive field equations for the higher
states using the approach of this paper.
It is interesting, however, that according to DMW's formula
Eq.~(\ref{eqn.dmw.pred}), the charges (such as $M$, $A$ and $B$) at
fixed $n-m$ are related.  Thus, for example, the charge of any state
with $n=m$ should be 
proportional to the mass of the blackhole at 
$n=1=m$.  Presumably this is a consequence of the $W_{\infty}$
symmetry in the string theory, but further investigation is beyond the
scope of this paper. 

\noindent{\large{\bf Acknowledgment}}

\noindent The author thanks Jim McCarthy with whom he had many
discussions regarding many points in this paper.

\noindent{\large{\bf References}}\newline
\inbib{gsw}{M.B. Green, J.H. Schwarz and E. Witten, {em Superstring Theory}
(Cambridge University Press, 1987), vol. I.}
\inbib{poly}{A.M. Polyakov, {\em Phys. Lett.} {\bf B103}, 207--210 (1981).}
\inbib{friedan}{D. Friedan, {\em Phys. Rev. Lett.} {\bf 45}, 1057--1060 (1980),
{\em Ann. Phys} {\bf 163}, 318--419 (1985).}
\inbib{cfmp}{C.G. Callan, D. Friedan, E.J. Martinec and M.J. Perry,
{\em Nucl. Phys.} {\bf B262}, 593--609 (1985).}
\inbib{ds}{S.R. Das and B. Sathiapalan, {\em Phys. Rev. Lett.} {\bf 56},
2664--2667 (1986).}
\inbib{bm}{T. Banks and E. Martinec, {\em Nucl. Phys.} {\bf B294},
733--746 (1987).} 
\inbib{jjj}{J. Hughes, J. Lu and J. Polchinski, {\em Nucl. Phys.} {\bf
B316}, 15--35 (1989).}
\inbib{gm}{P. Ginsparg and G. Moore, {\em Lectures on 2-D gravity and 2-D
string theory}, given at the TASI summer school (1992), hep-th/9304011.}
\inbib{polchrev}{J. Polchinski, {\em What is String Theory?}, Les
Houches Summer School, session 62 (1994), hep-th/9411028.}
\inbib{mp}{G. Moore and R Plesser, {\em Phys. Rev.} {\bf D46},
1730--1736 (1992).} 
\inbib{difk}{P. Di Franscesco and D. Kutasov, {\em Phys. Lett.} {\bf B261},
385--390 (1991).}
\inbib{polch}{J. Polchinski, {\em Nucl. Phys.} {\bf B362}, 125--140 (1991).}
\inbib{dmw}{A. Dhar, G. Mandal and S.P. Wadia, {\em Nucl. Phys.} {\bf
B454}, 541--560 (1995).}
\inbib{dmw2}{A. Dhar, {\em Nucl. Phys.} {\bf B507}, 277--291 (1997).}
\inbib{bny}{R. Brustein, D. Nemeschansky and S. Yankielowicz,
{\em Nucl. Phys.} {\bf B301}, 224-246 (1988).}
\inbib{dhp}{E. D'Hoker and D.H.Phong, {\em Phys. Rev.} {\bf D35},
3890--3901 (1987).} 
\inbib{rand}{S. Randjbar-Daemi, A. Salam, J. A. Strathdee,
{Int. J. Mod. Phys.} {\bf A2}, 667--693 (1987).}
\inbib{amt}{O. D. Andreev, R. R Metsaev and A. A. Tseytlin,
{\em Sov. J. Nucl. Phys.} {\bf 51}, 359--366 (1990).}
\inbib{wein}{S. Weinberg, {\em Phys. Lett.} {\bf B156}, 309--314 (1985).}
\inbib{bar1}{K. Bardakci and L.M. Bernardo, {\em Nucl. Phys.} {\bf B505},
463--496 (1997).}
\inbib{bar2}{K. Bardakci {\em String Field Equations from Generalized
Sigma Model II} hep-th/9710249.}
\inbib{buch}{I.L. Buchbinder, E.S. Fradkin, S.L. Lyakhovich and
V.D. Pershin, {\em Phys. Lett.} {\bf B304}, 239--248 (1993).}
\inbib{evans}{M. Evans, I. Giannakis and D.V. Nanopoulos, {\em
Phys. Rev.} {\bf D50}, 4022--4031 (1994); M. Evans and I. Giannakis, {\em
Phys. Lett.} {\bf B278}, 119--124 (1992).}

\end{document}